\documentclass[12pt]{article}
\usepackage{wider}

\begin{document}

\newcommand{\sqvb}{\ensuremath{ \langle \!\langle 0 |} }
\newcommand{\sqvk}{\ensuremath{ | 0 \rangle \!\rangle } }
\newcommand{\sqvn}{\ensuremath{ \langle \! \langle 0 |  0 \rangle \! \rangle} }
\newcommand{\wh}{\ensuremath{\widehat}}
\newcommand{\be}{\begin{equation}}
\newcommand{\ee}{\end{equation}}
\newcommand{\bea}{\begin{eqnarray}}
\newcommand{\eea}{\end{eqnarray}}
\newcommand{\ra}{\ensuremath{\rangle}}
\newcommand{\la}{\ensuremath{\langle}}
\newcommand{\rra}{\ensuremath{ \rangle \! \rangle }}
\newcommand{\lla}{\ensuremath{ \langle \! \langle }}
\newcommand{\str}{\rule[-.125cm]{0cm}{.5cm}}
\newcommand{\pr}{\ensuremath{^{\;\prime}}}
\newcommand{\ppr}{\ensuremath{^{\;\prime \prime}}}
\newcommand{\da}{\ensuremath{^\dag}}
\newcommand{\as}{^\ast}
\newcommand{\eps}{\ensuremath{\epsilon}}
\newcommand{\ve}{\ensuremath{\vec}}
\newcommand{\ka}{\kappa}
\newcommand{\non}{\ensuremath{\nonumber}}
\newcommand{\lf}{\ensuremath{\left}}
\newcommand{\rt}{\ensuremath{\right}}
\newcommand{\al}{\ensuremath{\alpha}}
\newcommand{\dfn}{\ensuremath{\equiv}}
\newcommand{\ga}{\ensuremath{\gamma}}
\newcommand{\ti}{\ensuremath{\tilde}}
\newcommand{\wti}{\ensuremath{\widetilde}}
\newcommand{\hs}{\ensuremath{\hspace*{.5cm}}}
\newcommand{\bet}{\ensuremath{\beta}}
\newcommand{\om}{\ensuremath{\omega}}
\newcommand{\kp}{\ensuremath{\kappa}}

\newcommand{\cO}{\ensuremath{{\cal O}}}
\newcommand{\cS}{\ensuremath{{\cal S}}}
\newcommand{\cF}{\ensuremath{{\cal F}}}
\newcommand{\cX}{\ensuremath{{\cal X}}}
\newcommand{\cZ}{\ensuremath{{\cal Z}}}
\newcommand{\cG}{\ensuremath{{\cal G}}}
\newcommand{\cR}{\ensuremath{{\cal R}}}
\newcommand{\cV}{\ensuremath{{\cal V}}}
\newcommand{\cC}{\ensuremath{{\cal C}}}
\newcommand{\cP}{\ensuremath{{\cal P}}}
\newcommand{\cH}{\ensuremath{{\cal H}}}
\newcommand{\cN}{\ensuremath{{\cal N}}}

\newcommand{\pup}{\ensuremath{^{(p)}}}
\newcommand{\prpr}{\ensuremath{\prime \prime }}

\newcommand{\hsp}{\ensuremath{\hspace*{5mm} }}
\newcommand{\sbp}{\ensuremath{_{[p]} }}

\newcommand{\xxx}[1]{}
\newcommand{\yyy}[1]{}
\newcommand{\zzz}[1]{}

\title{\bf 
Explanation, Evolution and Subjective Probability in Everett Quantum Mechanics with Positive Preclusion
\thanks{
This
work was sponsored by the Air Force under Air Force
Contract FA8721-05-C-0002. Opinions, interpretations,
conclusions, and recommendations are those
of the author and are not necessarily endorsed by
the U.S. Government.
}
}
\author{
Mark A. Rubin\\
Lincoln Laboratory\\ 
Massachusetts Institute of Technology\\  
244 Wood Street\\                         
Lexington, Massachusetts 02420-9185\\      
rubin@LL.mit.edu\\ 
}
\date{\mbox{}}
\maketitle

\begin{abstract}

The usual interpretational rule of quantum mechanics which states that outcomes do not occur  
when their weights are zero  is changed
so as to preclude outcomes with weights less than a small but  positive  value. With this ``positive preclusion'' rule, and in
the absence of any notion of objective probability, Everett quantum mechanics
has the explanatory power to account for the evolution of organisms with subjective expectations of probability that are in accord
with the Born rule. Positive preclusion also allows for the  derivation of a connection between weight and relative
frequency in situations involving a  finite number of measurements.

\mbox{}

\noindent Key words: Everett interpretation, probability, preclusion, evolution, Heisenberg picture

\end{abstract}

\section{Introduction} \label{SecIntro}
\xxx{SecIntro}

An Everett interpretation \cite{Everett57}-\cite{HewittHorsman09} of quantum mechanics in which subjective probability is present but objective 
probability\footnote{See, e.g., \cite{Saunders04} for a discussion of the relation between subjective
and objective probability.} (or some surrogate for
it) is absent is  incapable of explaining probabilistic outcomes \cite{Kent09}, most importantly the  biological evolution leading to organisms capable 
of making decisions or having subjective expectations such as ``likely'' or ``unlikely''   
in approximate  agreement with the Born rule---i.e., in agreement with experience,
something which we believe to be governed by quantum mechanics including the Born rule. 
In some branches of the wavefunction organisms with 
Born-rule-consistent subjective-probabilistic expectations will evolve and perceive that their expectations are often met; in 
other branches organisms will evolve with expectations different from the Born rule and perceive that their expectation are often
met; in others non-Born-rule-expecting organisms will be in constant states of surprise as their expectations are not met; and in
still others organisms expecting to find that the Born rule holds will be surprised on a regular basis. 
From our Born-rule-branch point of view the goings-on in these other non-Born branches are highly improbable, and ours highly probable.
But those in other branches would have different opinions on the matter, and without some objective  notion of probability we cannot 
claim that one opinion is correct and another is not.  
There is no sense in which one can explain why, in our branch, we expect Born-rule-consistent outcomes, and are unsurprised when our expectations are met.
The most one can say is ``This is how we are.''

One might be able to say more if one could  show, in a way not relying on any concept
of probability, that, e.g.,  evolution of non-Born-rule-expecting organisms is in some sense impossible, inherently contradictory. The decision-theoretic program initiated by Deutsch \cite{Deutsch99}-\cite{Wallace09} aims to show that, in a world (multiverse \cite{Deutsch01}) obeying the nonprobabilistic parts of quantum mechanics, an organism which obeys certain rules of rationality will act in accordance with the Born rule
(presumably from Born-rule-consistent expectations of outcome). Assume for the sake of argument that this connection is correct. Then essentially the same problem reemerges: How can we explain why we 
are in a branch in which organisms (at least from time to time) act rationally? Admittedly, the proposed rules of rationality 
are extremely natural, and it may indeed be possible to show that no organism capable of making decisions could evolve that did not in some sense incorporate 
these rules, and is therefore compelled to act in accordance with the Born rule (when acting rationally). If
that were shown it would constitute an ``anthropic'' \cite{Carter74}  argument: We are  organisms capable of making decisions, and any
such organism must incorporate the rules of rationality, hence make decisions (when they are made rationally) based on the Born rule.
But I do not believe that anything of this sort has yet been shown.\footnote{Zurek's approach \cite{Zurek03}, \cite{Zurek05} 
to probability in quantum mechanics
defines and quantifies probability based on the impossibility of exhibiting a preference between states related by the envariance symmetry. It thus falls into
the decision-theoretic category of approaches to probability and is also subject to this criticism.} 

The concept of subjective probability, i.e., probability  as an expectation, a cognitive state experienced by a living organism, has two great merits. It is a noncircular definition of probability, 
even when applied to the probability of the outcome of a single
experiment. And it is in accord with observation, both introspective and experimental (see following paragraph).
In any version of 
Everett quantum mechanics that provides a notion of explanation, the existence of evolution leading to organisms 
with a sense of subjective probability consistent with the Born rule
should follow from the theory as unambiguously as it does in orthodox quantum mechanics. There, objective probability is a
primitive concept not defined in terms of anything else and interacting with the Hilbert-space formalism of the nonprobabilistic
parts of quantum mechanics as a coequal partner. The Hilbert-space formalism determines the numerical value of the probability in
any given situation. And the probability concept, in conjunction with wavefunction collapse, gives the theory explanatory
power. Specifically, we do not observe organisms with subjective probabilistic expectations inconsistent with the Born rule because the evolution of
such organisms would be extreme improbable---and  the term ``improbable'' cannot be analyzed further.

I am of course not arguing that somewhere in the brains of living organisms are neural circuits equipped by evolution to
compute Hilbert-space inner products. What I am arguing is that subjective probability has as its  basis tendencies acquired
through biological evolution which cause the organism to act (and, if sufficiently high on the evolutionary ladder, to anticipate
and plan) in accordance with at least some features of the physical universe which, according to the Born rule, are highly probable.
For example, a plant turning its leaves to face the sun has ``learned,'' through evolution, that it is  likely (i.e., has greater
Born-rule weight) that in doing so it will increase
the number of photons of light that it absorbs. And recent experiments indicate that humans have an inborn tendency to learn by induction; that is,
to anticipate that the statistics found in past observations will persist in future observations (see, e.g., 
\cite{Kirkhametal02}-\cite{Turk-Brownetal08}).\footnote{``We report six 
experiments investigating whether 8-month-old infants are `intuitive statisticians.' \ldots  Our findings provide evidence that
infants possess a  powerful mechanism for inductive learning, either using heuristics or basic principles of
probability This ability to make inferences based on samples of information about the population develops early and
in the absence of schooling or explicit teaching.  Human infants may be rational learners from very early
in development \cite{XuGarcia08}.''}$\mbox{}^{,}$\/\footnote{Inductive learning cannot by itself constitute all that is involved in the
sense of subjective probability, as was implied in an earlier version of this paper. I am grateful to Rainer Plaga for pointing this 
out to me \cite{Plaga10}.}

The version of Everett quantum mechanics I present here  
explains  
biological evolution 
leading to Born-rule-consistent subjective expectation of probability without
invoking objective probability.  It
accomplishes this by  modifying an interpretational rule that is used,
explicitly or implicitly, in most interpretations of quantum mechanics:
namely, that outcomes with zero weight  are certain not to occur. This rule
is replaced with the idea, a version of which was first promoted in the context of the Everett interpretation
by Geroch \cite{Geroch84} and termed by him {\em preclusion}\/, that outcomes with {\em sufficiently small}\/ weights are
certain not to occur. The preclusion rule\footnote{Preclusion  rules identical or similar to the one used in the present paper have been  employed as key components of the single-outcome  interpretations of quantum mechanics  of
Sorkin \cite{Sorkin94}-\cite{Ghazi-TabatabaiWallden09b} and of Galvan \cite{Galvan07a}-\cite{Galvan08b}. Hanson  has employed  the related idea that low-weight worlds are ``mangled'' in his argument for objective probability from ``world counting'' in the Everett interpretation \cite{Hanson03}, \cite{Hanson06}. In classical probability theory the idea that sufficiently small probability is in some sense equivalent to certain nonoccurrence is
referred to as Cournot's principle \cite{Cournot1843}, \cite{Shafer07}.} plays the role, in Everett quantum mechanics, that objective probability
plays in orthodox quantum mechanics, in that it provides a notion of explanation for (at least) the results of biological evolution, in particular
the evolution of Born-rule-consistent subjective expectations of probability.

\section{Weight and zero preclusion}

The {\em weight}\/ associated with an event $E$\/ idealized as occurring at a moment of time $t$\/ is
\be
W_E(t)=\la \psi(t)|\wh{P}_E|\psi(t)\ra,
\label{WEtSP}
\ee
\xxx{WEtSP}
where $|\psi(t)\ra$\/ is the Schr\"{o}dinger-picture state at time $t$\/ and $\wh{P}_E$\/ is the Schr\"{o}dinger-picture projection operator corresponding to the event $E$\/.

In orthodox quantum mechanics (\ref{WEtSP}) is given meaning via the combined
concepts of probability and wavefunction reduction. Immediately after a measurement
with possible outcomes $E_i$\/, $i=1,2,\ldots,n$\/, one of these outcomes, say $E_j$\/, randomly occurs. The others, $E_i$\/ for $i\neq j$\/, do
not occur. Immediately after the measurement is made, the wavefunction is an eigenfunction of $\wh{P}_{E_j}$\/ with unit eigenvalue:
\be
\wh{P}_{E_j}|E_j\ra=|E_j\ra.
\label{PEjpsiEJ}
\ee
\xxx{PEjpsiEJ}
The  probability $P_{E_j}(t)$\/
that outcome $E_j$\/ occurs is equal to the weight $W_{E_j}(t)$\/ calculated via (\ref{WEtSP}):
\be
P_{E_j}(t)=W_{E_j}(t).
\label{PjteqEjt}
\ee
\xxx{PjteqEjt}
We may not be able to define this last statement
regarding probability---or, indeed, any statement regarding probability---in terms of anything more primitive, but we feel that we understand what it means. 

The idea of randomness or unpredictability, which is one part of the concept of probability \cite{Cramer46},
emerges naturally from the Everett interpretation. In general, there is no possibility of an observer Alice who is about to make 
a measurement at time $t$\/ predicting what ``she'' will see at time $t\pr$\/ immediately following $t$\/, because there will
in general not be a unique ``she'' in existence at   $t\pr$\/ but, rather, several ``shes,'' ``Alice-who-saw-outcome-$E_1$\/,''
``Alice-who-saw-outcome-$E_2$\/,'' etc.  But what meaning can we give to the quantitative idea of probability in the Everett interpretation?
What meaning can we attach to the value of $W_E(t)$\/?

Actually, there is a special case in which the meaning of $W_E(t)$\/ is completely unproblematic, in the Everett or any other interpretation of quantum mechanics; namely, the case in which
\be
W_{E}(t)=0.
\label{WEteq0}
\ee
\xxx{WEteq0}
In this case there is no component of the state vector $|\psi(t)\ra$\/ corresponding to outcome $E$\/. The meaning of the weight,
in this case is: After the measurement is made, Alice will {\em not}\/ see outcome $E$\/ (i.e., no Everett copy of Alice will
see $E$\/).

Furthermore, in the Everett interpretation, there is a meaning that can be attached to the complementary case
\be
W_{E}(t) > 0.
\label{WEtgt0}
\ee
\xxx{WEteq0}
In this case, after the measurement an Everett copy of an observer who has observed the event $E$\/ {\em will}\/ exist.\footnote{Of course there is another
special case, the case of certainty, $W_{E}(t)=1$\/,  in which the only Everett copy of the observer will observe $E$\/.  This case will not play a preferred role in what follows.} 

We can summarize this state of affairs by defining the {\em existence indicator function}\/ for event $E$\/ at time $t$\/,  $X_E(t)$\/:
\bea
X_{E}(t)&=&0, \;\;\; W_E(t)=0, \label{X0Et0}\\
      &=&1, \;\;\; W_E(t)> 0. \label{X0Et1}
\eea
\xxx{X0Et0,X0Et1}
\yyy{QNB3 p131}
The outcome $E$\/  does not occur  at time $t$\/, $X_E(t)=0$\/, if the weight is zero, and it occurs, $X_E(t)> 0$\/, if the
weight is nonzero. We will
refer to (\ref{X0Et0}), (\ref{X0Et1}) as the {\em zero preclusion rule}\/.

\section{The two-level ontology of Everett quantum mechanics in the Heisenberg picture}

In the Heisenberg picture of quantum mechanics it is the operators rather than the state vector that evolve
in time.
The elements of the formalism are a time-independent
state vector $|\psi_0\ra$\/ related to the time-dependent Schr\"{o}dinger-picture state vector
by
\be
|\psi_0\ra=\wh{U}\da(t)|\psi(t_0)\ra,
\ee
where $t_0$\/ is an initial time that can be chosen arbitrarily, and time-dependent operators $\wh{A}(t)$\/,  $\wh{B}(t)$\/, etc..
The operators evolve in time according to the rule
\be
\wh{A}(t)=\wh{U}\da(t)\; \wh{A} \;\wh{U}(t)
\label{At}
\ee
\xxx{At}
\yyy{QNB3 p134}
where $\wh{U}(t)$\/ is the unitary time-evolution operator, a function of other operators, satisfying
\be
\wh{U}(t_0)=1,
\label{Ut0}
\ee
\xxx{Ut0}
\yyy{QNB3 134}
and operators without explicit time arguments are Schr\"{o}dinger-picture
operators equal to their Heisenberg-picture counterparts at $t=t_0$\/:
\be
\wh{A}=\wh{A}(t_0),\;\;\;\wh{B}=\wh{B}(t_0),\;\;\;\mbox{\rm etc..}
\label{AeAt0}
\ee
\xxx{AeAt0}
\yyy{QNB3 p134}

The operators are defined at all times (or at all places and times, if they are operators in quantum field theory), and have
values at all times (e.g., they can be explicitly represented as infinite-dimensional matrices) regardless of the zero or nonzero
value obtained when computing a weight or any other matrix element. Particularly in the Deutsch-Hayden \cite{DeutschHayden00} version of the Heisenberg picture it is natural to think of the operators
as being ``real,'' as ``existing'' in the world.  For example, in a recent model \cite{Rubin09} of measurement in Deutsch-Hayden quantum field theory, the locality of the
theory depends on taking this viewpoint\footnote{The quantity which plays the role of the weight in \cite{Rubin09} is not an expectation value of a projection operator
but, rather, the expectation value of the integral of the field-theoretic number-density operator.}. 

If we accept the operators as real, then, in light of the zero preclusion rule, Everett quantum mechanics in the Heisenberg picture
has a two-level ontology. The operators are real at all times, but events, in particular observers' awareness of the outcomes
of measurements, are only real when the corresponding existence indicator functions, which obtain their values from the operators and state vector,\footnote{In the Deutsch-Hayden picture we can simply say ``obtain their values from the operators,'' since the state
vector is a purely conventional constant containing  no information.}
are nonzero. The connection between the operators and the reality, or lack thereof,
of objects is given by the existence indicator functions in a one-way manner, since operators affect the existence of objects but  operators
evolve according to equations of motion which are unaffected by the existence of objects.

\section{Positive preclusion, evolution, and subjective probability}\label{SecPP}

So, we can modify Everett quantum mechanics without in any way affecting the underlying equations of 
motion---equations which in a myriad of experiments have been born out to a high degree of accuracy---if 
we modify only the rule governing the value of the existence indicator function. Rather than the zero preclusion rule
(\ref{X0Et0}), (\ref{X0Et1}), we will take as a postulate the {\em positive preclusion rule}\/:
\bea
X_{\eps_P,E}(t)&=&0, \;\;\; W_E(t) \leq \eps_P, \label{XEt0}\\
      &=&1, \;\;\; W_E(t)> \eps_P,
\label{XEt1}
\eea
\xxx{XEt0,XEt1}
\yyy{QNB3 p131}
where the {\em preclusion parameter}\/ $\eps_P$\/ is a small positive number,
\be
0 < \eps_P \ll 1.\label{epsP}
\ee
\xxx{epsP}
The outcome $E$\/  does not occur  at time $t$\/, $X_{\eps_P,E}(t)=0$\/, if the weight is less than or equal to $\eps_P$\/, and it occurs, $X_{\eps_P,E}(t)=1$\/, if the
weight is greater than $\eps_P$\/.

There is no shortage of events of low probability which are known to have occurred.  
Therefore the preclusion parameter $\eps_P$\/ must
be extremely small, so as not to be in conflict with experience. It will therefore play almost no role in most events, in
and out of the physics laboratory.  Improbable events---winning the lottery, contracting a rare disease, etc.---will all exist
in the sense of (\ref{XEt1}), and to just the same degree as highly probable or even inevitable occurrences of the death-and-taxes variety.

The weights (\ref{WEtSP}) for all events will continue to be determined by the initial state vector
and the   equations of motion for the operators, and 
will have the same values that
they would have if we decided to interpret them as probabilities as in the orthodox interpretation. 
With only a preclusion rule at our disposal we cannot say which events among those which exist are ``probable'' or ``improbable.''
But, in any situation in which, in orthodox quantum mechanics, we would say that something is more or less probable, we {\em can}\/
say that the same thing has a greater or lesser weight. 

In particular, we can say that the biological evolution of an organism 
that learned\footnote{``Learning'' refers both to hard-wiring of the nervous system produced over many generations by evolution  as well as to the fruits of
the  ability, imbued in the organism by evolution, to learn from its own experience. E.g., visual systems  have been shown
to tune themselves, during the development of a given individual, to respond to features  found in the environment
to which the organism is exposed \cite{KandelSchwartz85}.} a sense of subjective probabilistic expectation different from the Born rule would
be of extremely low weight.  To see the correctness of this statement, replace ``weight'' by ``probability'' and consider that an organism
which could not learn an approximation to the Born rule 
would be at a disadvantage {\em vis a vis}\/ one which
could.   We therefore make the

\begin{quote}

\bf Born evolution assumption: \rm The weight for the evolution of a sentient organism which has subjective expectations of probability
significantly different from the Born rule 
is below the preclusion constant $\eps_P$\/. 
(The meaning of  ``significantly different'' is, of course,  ``different enough so as to drop the weight for the evolution of the organism
below $\eps_P$\/.'')

\end{quote}

The explanation of subjective probability is then simply:
Everett copies of organisms with subjective probability significantly different from the Born rule do not exist, since the weight for the event of
their biological evolution, after a sufficiently long time, will be less than $\eps_P$\/.

This is all that a theory of probability need account for. All the phenomena that are usually described
as related to, or examples of, probability ultimately reduce to the degree of surprise, or lack thereof, experienced
by a living organism at a turn of events.  What does it mean, e.g., that photons directed at a 50:50 beamsplitter
have an equal chance of being transmitted or reflected? Does it mean that in any group of photons exactly half will be transmitted?
Does it mean that it will never be the case that all of the photons is a group of 100 will be reflected? The only objective
fact about probability that can be described without using the term ``probability'' or a synonym is, in fact, the subjective experience of surprise at  certain events and lack of surprise at others.
 
There is of course a long chain of association from basic cognitive processes to the cognition, e.g., of a physicist who learns
quantum mechanics, calculates the cross-section for certain scattering process, gathers data in the laboratory and then is unsurprised
(surprised) that the data is close to (far from) the predicted outcome. Still, it is reasonable to ask if
we  can construct a model, even if extremely simplfied, of the preclusion of non-Born-expecting organisms. In fact, such a model already
exists; namely the quantum-mechanical model of the measurement of relative frequency (see Appendix). This  device observes the outcome 
of $N$\/ identical quantum-mechanical experiments and records the relative frequency of the a given outcome. In this sense it ``learns,''
say, the relative frequency to expect for ``spin up'' in a Stern-Gerlach experiment. For $N$\/ sufficiently large, the only
Everett copy of the device that will not be precluded is the one corresponding to the outcome with the largest Born
weight. Any organism or automaton that  bases its expectations on a relative frequency device which has been trained in this manner
will subsequently be more or less surprised at the outcome of a relative-frequency experiment on $N\pr \ll N$\/ systems to the degree
that the outcome is farther from or closer to the Born value.

\section{Summary and discussion}

Subjective expectation of probability is an observed fact of nature, but by itself does not allow us to explain probabilistic events.
Objective probability does, and in particular enables us to explain the evolution of subjective expectation of probability
consistent with objective probability. However, incorporating objective probability into the Everett interpretation is problematic;\footnote{See, e.g., \cite{Maudlin94}. However, this opinion is not universal; see \cite{Saunders98}.}
even in orthodox quantum mechanics it is an additional structure that exists alongside the Hilbert-space formalism.
Positive preclusion posits, within the deterministic logical structure of the Everett interpretation, the nonexistence of extremely low-weight events. In conjunction with the Born evolution assumption, it can therefore
explain why our subjective expectations of probability match the Born rule. It does so while adding to the theory only a minimal extension of
the notion of zero preclusion.  (Zero preclusion, in turn,  arguably flows directly from the formalism, since,  in all interpretations of quantum mechanics, absent components of the wavefunction correspond
to nonoccurring events.) 

The observation that  a theory of probability need only explain the evolution of subjective probability is
a critical one in the positive preclusion scheme. Competition acting over the
long time scales of biological evolution  causes species with evolutionary experience (and thus evolved behavior/expectation)
significantly at variance to the Born rule to be precluded, while the smallness of the preclusion constant ensures that surviving,
Born-rule-expecting, organisms will subsequently experience surprising low-weight as well as unsurprising high-weight events.

The internal consistency of the positive preclusion approach is assured by fact that the equations of motion are unaffected
by the change from zero preclusion.  
 However, it is certainly possible that Everett quantum mechanics with positive preclusion could be found to be
{\em inconsistent with experience}\/; it must be regarded as a new {\em theory}, not simply a new interpretation. 
Observations that would falsify the theory are ones that 
would be inconsistent with the Born evolution assumption. E.g., it is principle possible to compute the weight for
the evolution of various features of organisms \cite{Fisher30}-\cite{BlytheMcKane}; in particular, one could in principle compute the weight for the
evolution of a non-Born-expecting species. This would set a lower bound for $\eps_P$\/. If another event were then
observed to occur for which the weight was below this lower bound,  the theory would be falsified.\footnote{This assumes that $\eps_P$\/
is a fixed constant of nature.  An alternative  possibility is that $\eps_P$\/
emerges dynamically from the lower
level, e.g. by virtue of nonlinearities in the basic equations arising from quantum-gravitational effects \cite{Weissman99}, \cite{Weissman06}. In that case
$\eps_P$\/ might vary from situation to situation. In such a scenario it might be possible to relate the smallness of $\eps_P$\/ to
the weakness of the gravitional force relative to other forces.}

The  explanatory power which positive preclusion gives to the Everett interpretation is
limited compared to the explanatory power apparently afforded the orthodox interpretation by objective probability,
in that 
positive preclusion seems to have no place for what might be called ``normative probability.'' If someone were
to ask why it is that she is surprised at low-weight events, the positive preclusionist could answer: ``You evolved that way, because
Everett copies of you which are not surprised at such events  are of too low weight to have survived.'' But if
that same person were to ask why she {\em should}\/ be surprised at a low-weight event,  the positive preclusionist would have no answer. With objective
probability, on the other hand, we would of course simply say ``because it's improbable.''  But  it is not clear that
the inability of positive preclusion to provide this sort of explanation is  a deficiency. Rather, it may simply be the elimination 
of an illusion which, like absolute simultaneity,  has no objective correlate in the physical world.

\renewcommand{\theequation}{A-\arabic{equation}}
  \setcounter{equation}{0}  

\section*{Appendix: Relative frequency and positive preclusion}

Starting from the zero preclusion rule,\footnote{Or in some cases from the rule $W_E(t)=1 \Rightarrow$\/ certain occurrence.} many derivations \cite{Finkelstein65}-{\cite{Rubin03} have been given of  probability in the sense
of {\em relative frequency in the limit of an infinite number of identically-prepared measurement situations}\/.  
The mathematics involved in considering an infinite
number of measurements has been questioned \cite{CassinelloSanchezGomez96},\cite{CavesSchack05} and defended 
\cite{VanWesep06},\cite{Landsman08}. 
If we 
start from the positive preclusion rule we can apply essentially the same calculational steps as in these previous derivations to 
prove, in a straightforward and rigorous fashion, a relative-frequency theorem 
which involves only a finite number of measurements.

Rather than taking the most general case of measurements involving an arbitrary discrete or continuous number of outcomes,
we consider only measurements of two-state systems (qubits). Then the result for the infinite-number-of-copies limit
in \cite{Rubin03} enables us to prove immediately the

\noindent {\bf Positive preclusion relative frequency theorem:} Let $N$\/ observers measure $N$\/ two-state systems,
such that each of the systems is measured by one of the observers and such that all of the systems are prepared in the same quantum state
\be
|\psi (c_1,c_2)\ra=c_1|1\ra + c_2|2\ra,
\label{psic1c2}
\ee
\xxx{psic1c2}
where $|c_1|^2+|c_2|^2=1$\/. Subsequent to these $N$\/  measurements of systems by observers, let a single additional ``relative-frequency observer''
measure all $N$\/ of the initial observers to determine the relative frequency of the result ``1,'' i.e. the fraction of the $N$\/ observers who determined that the two-level system they measured was in state $|1\ra$\/. If the relative-frequency observer is of finite size and energy, and therefore can only record  the results of measurements to a finite precision (so that there are only a finite number of possible outcomes
to the relative frequency measurement), then there is a number $N_B$\/ such that if
$N > N_B$\/ there will only exist  one Everett copy of the relative-frequency
observer subsequent to the relative-frequency measurement. This copy will have observed the Born-rule relative frequency $|c_1|^2$\/ to within the precision of which it is capable.

\noindent {\bf Proof:} I have previously shown, in \cite{Rubin03}, that in the limit $N\rightarrow\infty$\/, the weight for the  Everett copy
of the relative-frequency observer which perceives relative frequency closest  to the Born-rule value will remain nonzero (in fact, approach unity), while the weights for those which perceive relative frequencies farther from the Born-rule value will approach zero. But of course to say that $\lim_{N\rightarrow\infty} W=0$\/ is simply to say that
for any positive real number $\eps_P$\/, no matter how small, there is a number $N_B$\/ such that, for $N > N_B$\/,
$|W| < \eps_P$\/. From the positive preclusion rule (\ref{XEt0}), (\ref{XEt1}) we conclude that for $N > N_B$\/ there will
only be one surviving Everett copy of the relative frequency observer, the one which has obtained a relative-frequency measurement
as close as possible to the Born-rule value.\footnote{If two possible measured relative frequencies are equally  close to the Born-rule value then the Everett copies corresponding to each of these will survive. See \cite{Rubin03}. }  {\bf Q.E.D.}

For details of the calculation, the relative-frequency measurement model, and the ontology of the Everett interpretation in the Heisenberg    
   picture, see \cite{Rubin03}.

\section*{Acknowledgments}

I would like to thank Jianbin Mao and Allen J. Tino for helpful discussions.
I am particularly grateful to  Michael Weissman for  pointing out to me 
the possibility of modifying the rule linking mathematical formalism to probability while retaining unchanged
the structure of the underlying formalism, and to Rainer Plaga for extensive comments on an earlier version of this paper.


\begin{thebibliography}{99}

\bibitem{Everett57}H. Everett~III, `` `Relative state' formulation of quantum 
mechanics,'' {\em Rev. Mod. Phys.}\/ {\bf 29}  454-462 (1957). Reprinted in \cite{DeWittGraham73}.

\bibitem{DeWittGraham73}B.~S.~DeWitt and
N.~Graham, eds., {\em The Many Worlds Interpretation of Quantum Mechanics}
(Princeton University Press, Princeton, NJ, 1973). 

\bibitem{Price95}  M.~C.~Price, ``The Everett FAQ,'' http://www.hedweb.com/manworld.htm  (1995). 

\bibitem{Vaidman02} L.~Vaidman, ``The many-worlds interpretation of quantum mechanics,'' Stanford Encyclopedia of Philosophy,
http://plato.stanford.edu/entries/qm-manyworlds/(2002).

\bibitem{HewittHorsman09}C.~Hewitt-Horsman, ``An introduction to many worlds in quantum computation,'' {\em Found. Phys.}\/ {\bf 39},
826-902 (2009); arXiv:0802.2504.

\bibitem{Saunders04}S.~Saunders, ``What is probability?,''quant-ph/0412194.

\bibitem{Kent09}A.~Kent, ``One world versus many: the inadequacy of Everettian accounts of evolution, probability, and scientific confirmation,''
to appear in S.~Saunders, J.~Barrett, A.~Kent and D.~Wallace, eds.,{\em Many Worlds? Everett, Quantum Theory and Reality}\/,   (Oxford University Press);
quant-ph/0905.0624. 


\bibitem{Deutsch99}D.~Deutsch, ``Quantum theory of probability and decisions,'' {\em Proc.
Roy. Soc. Lond. A}\/ {\bf 455}, 3129-3137, (1999); quant-ph/9906015.

\bibitem{DeWitt98}B.~DeWitt, ``The quantum mechanics of isolated systems,'' {\em Int. J. Mod. Phys. A}\/ {\bf 13}, 1881-1916 (1998).

\bibitem{Polley99} L.~Polley, ``Quantum-mechanical probability from the symmetries of two-state systems,''   quant-ph/9906124. 


\bibitem{Polley01}L.~Polley, ``Position eigenstates and the statistical axiom of quantum mechanics,'' in {\em Foundations of Probability and Physics, Vaxjo, Nov 27 - Dec 1, 2000}\/;
 quant-ph/0102113.

\bibitem{Wallace03} D.~Wallace, ``Everettian rationality: defending Deutsch's approach
to probability in the Everett interpretation,'' {\em Stud. 
Hist. Phil. Mod. Phys.}\/ {\bf 34}, 415-439 (2003); quant-ph/0303050.

\bibitem{Wallace06} D.~Wallace, ``Epistemology quantized: circumstances in which we
should come to believe in the Everett interpretation,'' forthcoming in
{\em Brit. J. Phil. Sci}\/; 
http://philsci-archive.pitt.edu.

\bibitem{Wallace07}D.~Wallace, ``Quantum probability from subjective likelihood: Improving
on Deutsch's proof of the probability rule,''{\em Stud. 
Hist. Phil. Mod. Phys.}\/  {\bf 38}, 311-332 (2007); quant-ph/0312157.

\bibitem{Wallace09}D.~Wallace, ``A formal proof of the Born rule from decision-theoretic assumptions,''
to appear (under the title ``How to prove the Born rule'') in S.~Saunders, J.~Barrett, A.~Kent and D.~Wallace, eds.,{\em Many Worlds? Everett, Quantum Theory and Reality}\/,   (Oxford University Press); quant-ph/0906.2718.

\bibitem{Deutsch01}D.~Deutsch, ``The structure of the multiverse,'' {\em Proc. R. Soc. Lond. A} {\bf 458} 2911-2923 (2002);
quant-ph/0104033.

\bibitem{Carter74}B.~Carter, ``Large number coincidences and the anthropic principle in cosmology,''
 in 
{\em IAU Symposium 63: Confrontation of Cosmological Theories with Observational Data}, (Reidel, Dordrecht, 1974) 291-298.


\bibitem{Zurek03}W.~H.~Zurek, ``Environment-assisted invariance, causality, and probabilities in quantum physics,''
{\em  Phys. Rev. Lett.}\/ {\bf 90}, 120403 (2003); quant-ph/0211037.


\bibitem{Zurek05}W. H. Zurek, ``Probabilities from entanglement, Born's rule from envariance,'' 
{\em Phys. Rev. A}\/ {\bf 71}, 052105 (2005); quant-ph/0405161.


\bibitem{Kirkhametal02}N.~Z.~Kirkham, J.~A.~Slemmer and S.~P.~Johnson, ``Visual statistical learning in infancy:
evidence for a domain general learning mechanism,'' {\em Cognition}\/ {\bf 83} B35-B42 (2002).

\bibitem{XuGarcia08} F.~Xu and V.~Garcia, ``Intuitive statistics by 8-month-old infants,'' {\em Proc. Nat. Acad. Sci.}\/
{\bf 105}, 5012-5015 (2008).

\bibitem{Turk-Brownetal08}N.~Turk-Browne, B.~J.~Scholl, M.~M.~Chun and M.~K.~Johnson, ``Neural evidence of statistical learning:
Efficient detection of visual regularities without awareness,'' {\em J. Cog. Neurosci.}\/ {\bf 21}, 1934-1945 (2008).



\bibitem{Plaga10}R.~Plaga, personal communication.

\bibitem{Geroch84}R.~Geroch, ``The Everett interpretation,'' {\em No\^{u}s}\/ {\bf 18}, 617-633 (1984).


\bibitem{Sorkin94} R.~D.~Sorkin, ``Quantum mechanics as quantum measure theory,'' {\em Mod. Phys. Lett.}\/ {\bf A9}  3119-3128 (1994); gr-qc/9401003.

\bibitem{Sorkin97} R.~D.~Sorkin, ``Quantum measure theory and its interpretation,'' in  D.~H.~Feng and B.-L.~Hu, eds., 
{\em Quantum Classical Correspondence: Proceedings of the 4th Drexel Symposium on Quantum Nonintegrability, held Philadelphia, September 8-11, 1994}\/, 229-251 (International Press, Cambridge Mass., 1997); gr-qc/9507057.

\bibitem{Sorkin07a} R.~D.~Sorkin, ``An exercise in `anhomomorphic logic' ,''
{\em J. Phys.}\/ Conf. Ser. {\bf 67}, 012018 (2007); quant-ph/0703276.

\bibitem{Sorkin07b} R.~D.~Sorkin, ``Quantum dynamics without the wave function,'' {\em J. Phys. A} {\bf 40}, 3207-3222 (2007); quant-ph/0610204.


\bibitem{DowkerGhaziTabatabai08b} F.~Dowker and Y.~Ghazi-Tabatabai, ``Dynamical wave function collapse models in quantum measure theory,''
 {\em J. Phys. A}\/ {\bf 41}, 205306 (2008); arXiv:0712.2924.

\bibitem{DowkerGhaziTabatabai08a} F.~Dowker and Y.~Ghazi-Tabatabai, ``The Kochen-Specker theorem revisited in quantum measure theory,'' 
{\em J. Phys. A}\/ {\bf  41}, 105301 (2008); arXiv:0711.0894.

\bibitem{Barnettetal07}M.~Barnett, F.~Dowker and D.~Rideout, ``Popescu-Rohrlich boxes in quantum measure theory,''
  {\em J. Phys. A} {\bf 40},  7255-7264 (2007); quant-ph/0605253.

\bibitem{Craigetal07}D.~Craig, F.~Dowker, J.~Henson, S.~Major, D.~Rideout and R.~D.~Sorkin, ``A Bell inequality analog in quantum measure theory,''
{\em  J. Phys. A}\/ {\bf 40}, 501-523 (2007);  quant-ph/0605008.


\bibitem{Ghazi-TabatabaiWallden09a}Y.~Ghazi-Tabatabai and P.~Wallden,
``Dynamics \& predictions in the co-event interpretation,''
     {\em J. Phys. A}\/, {\bf 42}, 235303 (2009); arXiv:0901.3675.

\bibitem{Ghazi-Tabatabai09}Y.~Ghazi-Tabatabai, ``Quantum measure theory: A new interpretation,''arXiv:0906.029.
 
\bibitem{Ghazi-TabatabaiWallden09b}Y.~Ghazi-Tabatabai and P.~Wallden,
 ``The emergence of probabilities in anhomomorphic logic,''
 {\em J. Phys.}\/ Conf. Ser. {\bf 174}, 012054 (2009);  arXiv:0907.0754. 


\bibitem{Galvan07a} B.~Galvan, ``Typicality vs. probability in trajectory-based formulations of quantum mechanics,''
{\em Found. Phys.} {\bf 37}, 1540-1562 (2007); arXiv:quant-ph/0605162.

\bibitem{Galvan07b} B.~Galvan, ``Origin of which-way information and generalization of the Born rule,'' 
arXiv:0705.2877.


\bibitem{Galvan08a} B.~Galvan,  ``Quantum mechanics and imprecise probability,''  {\em J. Stat. Phys.}\/ {\bf 131}, 1155-1167
(2008); arXiv:0711.3000.



\bibitem{Galvan08b} B.~Galvan, ``Generalization of the Born rule,''
{\em  Phys. Rev. A}\/ {\bf 78}, 042113 (2008); arXiv:0806.4935.


\bibitem{Hanson03} 
  R.~D.~Hanson, ``When worlds collide: Quantum probability from observer selection?,''
{\em Found. Phys.} {\bf 33}, 1129-1150 (2003);  quant-ph/0108070.  

\bibitem{Hanson06}R.~D.~Hanson, ``Drift-diffusion in mangled worlds quantum mechanics,''
{\em Proc. Roy. Soc. A}\/, {\bf 462},  1619-1627 (2006); quant-ph/0303114.






\bibitem{Cournot1843} A.~A.~Cournot, {\em Exposition de la th\'{e}orie des chances et des probabilit\'{e}s}\/  (Hachette,
Paris, 1843). Reprinted as Volume I of B.~Bru ed., {\em Antoine-Augustin Cournot, Oeuvres
compl\`{e}tes}\/ (Vrin, Paris, 1984).

\bibitem{Shafer07} G.~Shafer, ``From Cournot's principle to market efficiency,'' in J.~P.~Touffut ed. {\em Augustin
Cournot: Modelling Economics}\/ (Edward Elgar, 2007); http://www.glennshafer.com.


\bibitem{Cramer46}H.~Cram\'{e}r, {\em Mathematical Methods of Statistics}\/ (Princeton University Press, Princeton, 1946).

\bibitem{DeutschHayden00} D.~Deutsch and P.~Hayden,   ``Information flow in entangled
quantum systems,'' {\em Proc. R. Soc. Lond. A}\/ {\bf456} 1759-1774 (2000);
quant-ph/9906007.

\bibitem{Rubin09}   M.~A.~Rubin, ``Observers and locality in Everett quantum field theory,'' arXiv:0909.2673.
 
\bibitem{KandelSchwartz85} E.~R.~Kandel and J.~H.~Schwartz, {\em Principles of Neural Science}\/ (Elsevier, New York, 1985), p. 759.

\bibitem{Maudlin94}T.~Maudlin, {\em Quantum Non-Locality and Relativity,}\/ (Blackwell, Oxford, 1994).

\bibitem{Saunders98}S.~Saunders, ``Time, quantum mechanics and probability,'' {\em Synthese}\/ {\bf 114}, 373-404 (1998).

\bibitem{Fisher30} R.~A.~Fisher, {\em The Genetical Theory of Natural Selection}\/, (Clarendon Press, Oxford, 1930).
\bibitem{Wright31}S.~Wright, ``Evolution in mendelian populations,'' {\em Genetics}\/ {\bf 16}, 97-159 (1931).
\bibitem{Haldane32} J.~B.~S.~Haldane, {\em The Causes of Evolution} (Longmans Green \& Co., London, 1932).

\bibitem{Drossel}B.~Drossel, ``Biological evolution and statistical physics, {\em Adv. Phys.}\/ {\bf 50}, 209-295 (2001).

\bibitem{LassigValleriani02}M.~L\"{a}ssig and A.~Valleriani, eds.,  {\em Biological Evolution and Statistical Physics}\/ (Springer, Berlin, 2002).

\bibitem{Fisher07}D.~S.~Fisher, ``Evolutionary dynamics,'' in J.~P.~Bouchaud, M.~M\'{e}zard and J.~Dalibard, eds., {\em Complex
Systems}\/ (Elsevier, Amsterdam, 2007).

\bibitem{BlytheMcKane} R.~A.~Blythe and A.~J.~McKane, ``Stochastic models of evolution in genetics, ecology and linguistics,'' {\em J.
Stat. Mech.: Theory Exp.}\/,  {\bf P07018} (2007).


\bibitem{Weissman99}M.~B.~Weissman, ``Emergent measure-dependent probabilities from modified quantum dynamics without state-vector reduction,''
{\em Found. Phys. Lett.}\/ {\bf 12} (1999) 407-426; quant-ph/9906127.

\bibitem{Weissman06} M.~B.~Weissman, ``Irreversibility in collapse-free quantum dynamics and the second law of thermodynamics,'' quant-ph/0605031.

\bibitem{Finkelstein65}D.~Finkelstein, ``The logic of quantum mechanics,'' {\em Trans. N. Y. Aca. Sci.}\/ {\bf 25}, 621-637
(1965).

\bibitem{Hartle68}J.~Hartle, ``Quantum mechanics of individual systems,'' {\em Am. J. Phys.}\/ {\bf 36}, 704-712
(1968).

\bibitem{DeWitt72}B.~S.~DeWitt, ``The many-universes interpretation of quantum mechanics,'' in {\em Proceedings
of the International School of Physics `Enrico Fermi' Course IL: Foundations of
Quantum Mechanics}\/ (Academic Press, Inc., New York, 1972). Reprinted in 
\cite{DeWittGraham73}.

\bibitem{Graham73}N.~Graham, ``The measurement of relative frequency,'' in \cite{DeWittGraham73}.

\bibitem{Farhietal89}E.~Farhi, J.~Goldstone, and S.~Gutmann, ``How probability arises in quantum mechanics,''
{\em Ann. Phys.}\/ (NY) {\bf 192}, 368-382 (1989).

\bibitem{Okhuwa93}Y. Okhuwa, ``Decoherence functional and probability interpretation,'' {\em Phys. Rev.
D}\/ {\bf 48}, 1781-1784 (1993).

\bibitem{Rubin03}M.~A.~Rubin, ``Relative frequency and probability in the Everett interpretation of Heisenberg-picture quantum mechanics,'' {\em Found. Phys.} {\bf 33}  379-405 (2003); quant-ph/0209055.

\bibitem{CassinelloSanchezGomez96}A.~Cassinello and J.~L.~S\'{a}nchez-G\'{o}mez, ``On the probabilistic postulate of
quantum mechanics,'' {\em Found. Phys.}\/, {\bf 26}, 1357–1374 (1996).

\bibitem{CavesSchack05}C.~M.~Caves and R.~Schack, ``Properties of the frequency operator do not imply
the quantum probability postulate,'' {\em Ann. Phys.}\/ (NY) {\bf 315}, 123–146 (2005).

\bibitem{VanWesep06} R.~A.~Van~Wesep, ``Many worlds and the emergence of probability in quantum mechanics,''
{\em Ann. Phys.}\/ (NY) {\bf 321},2438-2452  (2006) ; quant-ph/0506024.

\bibitem{Landsman08}N~P.~Landsman, ``Macroscopic observables and the Born rule,'' arXiv:0804.4849.
    


\end{thebibliography}
\end{document}